# Receiver Resonant Frequency Adaptive Tracking in Wireless Power Transfer Systems Using Primary Variable Capacitor

Chang Liu, *Student Member, IEEE*, Wei Han, *Member, IEEE,* Guangyu Yan, *Student Member, IEEE,* Bowang Zhang, *Student Member, IEEE,* Chunlin Li, *Student Member, IEEE*

***Abstract*—Parameter variations within the resonant network of wireless power transfer (WPT) systems can cause drift in the resonant frequency, leading to a detuned system that requires higher power capacity and experiences reduced transfer efficiency. To address this issue, this paper presents an adaptive online receiver resonant frequency tracking scheme based solely on primary-side detection. The proposed method effectively compensates for parameter fluctuations in both primary and secondary resonators. The core of this approach is a switch-controlled capacitor (SCC) with a control angle calibrated during a system self-check process prior to high-power charging. Additionally, a two-step perturb-and-observe algorithm has been developed to perform online tracking while minimizing disturbances to the output power. Post-tracking, zero-voltage switching (ZVS) conditions can be achieved within a specified detuning range. To validate the efficacy of the proposed system, a 200W experimental platform was constructed. The measured results demonstrate that resonance is consistently maintained within the 79-90 kHz frequency range, as specified by the SAE J2954 standard. The maximum frequency tracking error and efficiency increase are 0.7 kHz and 9%, respectively. Notably, the tracking process is completed in less than 1 ms.**

## I. INTRODUCTION

WIRELESS power transfer (WPT), powering devices in a cordless way, exhibits the characteristics of higher safety, reliability and convenience. It has been increasingly used in electric vehicles [1], portable consumer electronics [2], medical implants, home appliances and other industrial applications [3], [4], [5]. The high efficiency of WPT systems stems from the elevated quality factor of the resonant network [6]. However, deviations from ideal resonance parameters inevitably lead to decreased system efficiency and output power. In practical scenarios, detuning arises due to several factors. Firstly, inherent inaccuracies in commercial capacitors coupled with discrete capacitance values preclude precise tuning, exacerbating system costs if improved accuracy is pursued. Secondly, ambient temperature variations and aging contribute to capacitance parameter shifts, thereby altering system resonant frequencies during operation [7], [8]. Thirdly, the mobile nature of receiving ends in WPT systems, compounded by magnetic material effects like ferrite, induces changes in coil self-inductance [9], aggravating system detuning. Finally, prevalent wireless charging standards like Qi for electronic devices (110kHz~205kHz) or SAE J2954 for electric vehicles (79kHz~90kHz) mandate a broad resonant frequency range. Consequently, receivers operating within this specified range are deemed compliant, necessitating transmitters to adapt to varying receiver resonant frequencies to enhance overall compatibility. Therefore, researching how to address the frequency mismatch issue in WPT systems is of paramount importance.

Research efforts can be categorized into three main approaches to address the detuning issues in WPT systems. The first approach focuses on parameter identification of the receiver before high-power transmission. For instance, optimization algorithms are employed in [10] for extensive frequency sweeps to identify the secondary resonant parameters and pure resistive loads. In [11], an auxiliary coil is used to sense the secondary current to identify the receiver's resonant frequency. However, due to the absence of corresponding reactive compensation circuits, these methods cannot restore the system to a dual-resonant state after identifying detuning.

The second approach addresses the efficiency reduction caused by detuning during normal high-power operation using control methods known as frequency tuning. Some studies, such as [9] and [13], employ primary-side frequency tuning by tracking the phase alignment of primary voltage and current to mitigate frequency mismatch. Other studies introduce different reactive compensation circuits on the secondary side, such as the active single-phase rectifier (ASPR) in [13], [14] and the variable reactor in [15]. However, these solutions cannot avoid issues arising from primary-side detuning. Consequently, in [9], [16], both primary and secondary sides are equipped with switch-controlled capacitors (SCCs) to adjust reactive power and restore the system from detuning at a fixed frequency. However, this significantly increases the system cost, volume, and control complexity of the secondary side. An alternative method incorporating an auxiliary coil and a decoupling transformer can realize capacitance tuning on both sides [17].

The third approach involves tracking the secondary-side resonant frequency and compensating the primary-side reactive power. Given that the resonant state of the secondary side has the most significant impact on system efficiency [18], after ensuring that the secondary resonant frequency aligns with the standard frequency band, this approach can enhance system efficiency under detuning conditions and reduce current stress on the primary side. For example, in [19], introducing an SCC allows multiple primary resonant frequencies for system parameter identification. After identification, the SCC control angle can be adjusted to achieve corresponding primary-side reactive compensation.

Table I presents a comparison of current methods for secondary resonant frequency identification or frequency tuning in WPT systems, where $L_p$ ($L_s$) and $R_p$ ($R_s$) represent primary (secondary) self-inductance and parasitic resistance of coupled coils, $C_p$ and $C_s$ are corresponding compensation capacitors, $M$ is the mutual inductance, and $R_L$ is the load resistance.

TABLE I
COMPARISON OF DIFFERENT APPROACHES IN THE EXISTING LITERATURE

| research | Identification/ Tuning time | Known Parameters | Secondary resonance | ZVS/ZPA |
|---|---|---|---|---|
| [10] | >33.77s | $L_p$, $C_p$, $R_p$ | No | No |
| [11] | 5-9ms | No | No | No |
| [12] | >500ms | $L_p$,$C_p$,$L_s$,$C_s$,$M$ | No | Yes |
| [14] | >1.4s | $L_p$,$C_p$,$L_s$,$R_{aux}$ | Yes | Yes |
| [16] | >600ms | No | Yes | Yes |
| [17] | 2ms | $M$,$R_L$ | Yes | Yes |
| [19] | 0.61s | $L_p$,$C_p$,$R_p$ | Yes | Yes |
| This work | 1ms | No | Yes | Yes |

It is evident that optimization algorithm-based approaches are time-consuming for parameter identification methods, require purely resistive loads, and are unsuitable for online applications [10]. Identification methods based on other control techniques necessitate auxiliary coils or active rectifiers [11][14]. For frequency tuning methods: primary frequency tuning cannot accommodate a wide range of detuning [12], while secondary or double-side frequency tuning methods require additional reactive compensation circuits on the secondary side [14][16][17], increasing the volume and cost of the receiver, making them unsuitable for universal transmitter designs. Moreover, most existing methods require knowledge of resonant parameters, mutual inductance or load information. This affects the system's generality, making these methods potentially ineffective under different operating conditions or requiring more time to calibrate system parameters. Therefore, this paper proposes an adaptive approach with the use of an SCC to achieve both receiver resonant frequency identification and primary-side reactance compensation, which has the following features:

1) Fast identification speed, allowing both primary and secondary sides to resonate simultaneously at the identified receiver resonant frequency.

2) The algorithm does not rely on known system parameters. System self-check mitigates the impact of parameter fluctuations within a specific range of primary resonators. Additionally, since the receiver is placed in the charging position during self-check, fluctuations in the transmitter's self-inductance by the presence of the receiver are accounted for.

3) No wireless communication or additional secondary-side circuitry is required.

4) Minimal system output variation during identification, with little impact on the load, making it suitable for online identification.

The remainder of this article is structured as follows: In Section II, we compare and analyze the characteristics of the SS-WPT system under varying detuning conditions. Subsequently, Section III comprehensively introduces the principle of the proposed method. Section IV presents experimental verifications by showing the frequency identification and tuning results under various conditions. Finally, Section V concludes the paper.

## II. DETUNING CONDITION ANALYSIS

### A. Characteristics of SS-WPT system

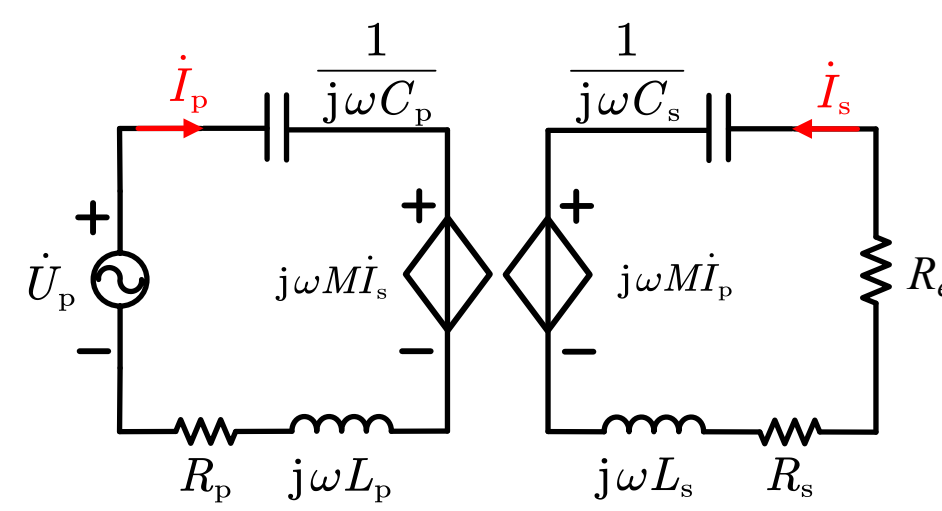

**Fig. 1.** Circuit model of SS-WPT system.

The SS-WPT possesses features of simplicity and load-independent resonant frequency while it is susceptible to parameter variations. As illustrated in Fig. 1., $R_e$ represents the equivalent load resistance. Therefore, the inherent resonant angular frequencies on primary and secondary sides are $\omega_p$ and $\omega_s$ as given by:

$$\begin{cases} \omega_p = 1/\sqrt{L_p C_p} \\ \omega_s = 1/\sqrt{L_s C_s} \end{cases} \tag{1}$$

The system general equation can be deduced in (2) based on Kirchhoff's Voltage Law (KVL) and the first harmonic analysis. Besides, $X_p$ and $X_s$ refer to the reactance of resonators on both sides, as given in (3).

$$\begin{bmatrix} \dot{U}_p \\ 0 \end{bmatrix} = \begin{bmatrix} R_p + jX_p & j\omega M \\ j\omega M & R_s + R_e + jX_s \end{bmatrix} \begin{bmatrix} \dot{I}_p \\ \dot{I}_s \end{bmatrix} \tag{2}$$

$$\begin{cases} X_p = \omega L_p - 1/(\omega C_p) \\ X_s = \omega L_s - 1/(\omega C_s) \end{cases} \tag{3}$$

Therefore, the transfer efficiency $\eta$ and input impedance $Z_{in}$ of the WPT system can be obtained as

$$\eta = \frac{\omega^2 M^2 R_L}{[(R_s + R_L)^2 + X_s{}^2] R_p + \omega^2 M^2 (R_s + R_L)} \tag{4}$$

$$Z_{in} = (R_p + \frac{(\omega M)^2 (R_s + R_e)}{(R_s + R_e)^2 + X_s^2}) + j(X_p - \frac{(\omega M)^2 X_s}{(R_s + R_e)^2 + X_s^2}) \tag{5}$$

It can be observed that the transfer efficiency can be maximized when secondary reactance $X_s$ is zero. The unit power factor can be achieved if $X_p$ is zero simultaneously. However, parameter variations, such as changes in the secondary capacitor, prevent $X_s$ from zero, resulting in frequency drift, reduced efficiency and power factor. Critical parameters for calculations are shown in Table II to analyze the detuning problem of the WPT system.

TABLE II
PARAMETERS FOR THEORETICAL CALCULATIONS

| Items | Symbol | Value | Unit |
|---|---|---|---|
| Input voltage RMS | $U_p$ | 40 | V |
| Primary-side capacitor | $C_p$ | 29.46 | nF |
| Secondary-side capacitor | $C_{s0}$ | 38.01 | nF |
| ESR of the transmitter coil | $R_p$ | 0.3 | Ω |
| Self-inductance of transmitter | $L_p$ | 118.3 | uH |
| ESR of the receiver coil | $R_s$ | 0.3 | Ω |
| Self-inductance of receiver | $L_s$ | 91.8 | uH |
| Coupling coefficient | $k$ | 0.1~0.2 | - |
| Equivalent load resistance | $R_e$ | 2~20 | Ω |
| Deviation coefficient of $C_{s0}$ | Δ | -0.2~0.2 | - |

*B. Analysis of Power Factor and Transfer Efficiency in detuned WPT system*

To comply with the SAE J2954 standard, the value of $C_s$ is initially designed to resonate with $L_s$ at 85 kHz, denoted as $C_{s0}$. When the secondary capacitance deviation occurs, the actual capacitance becomes $C_{s0}+\Delta\times C_{s0}$. Fig. 2 depicts the power factor and transfer efficiency while considering the detuning of the secondary-side system and utilizing the parameters outlined in Table II. It can be observed that as the deviation coefficient increases, the transfer efficiency $\eta$ and power factor decrease. Furthermore, the efficiency variation is more significant under weak coupling conditions.

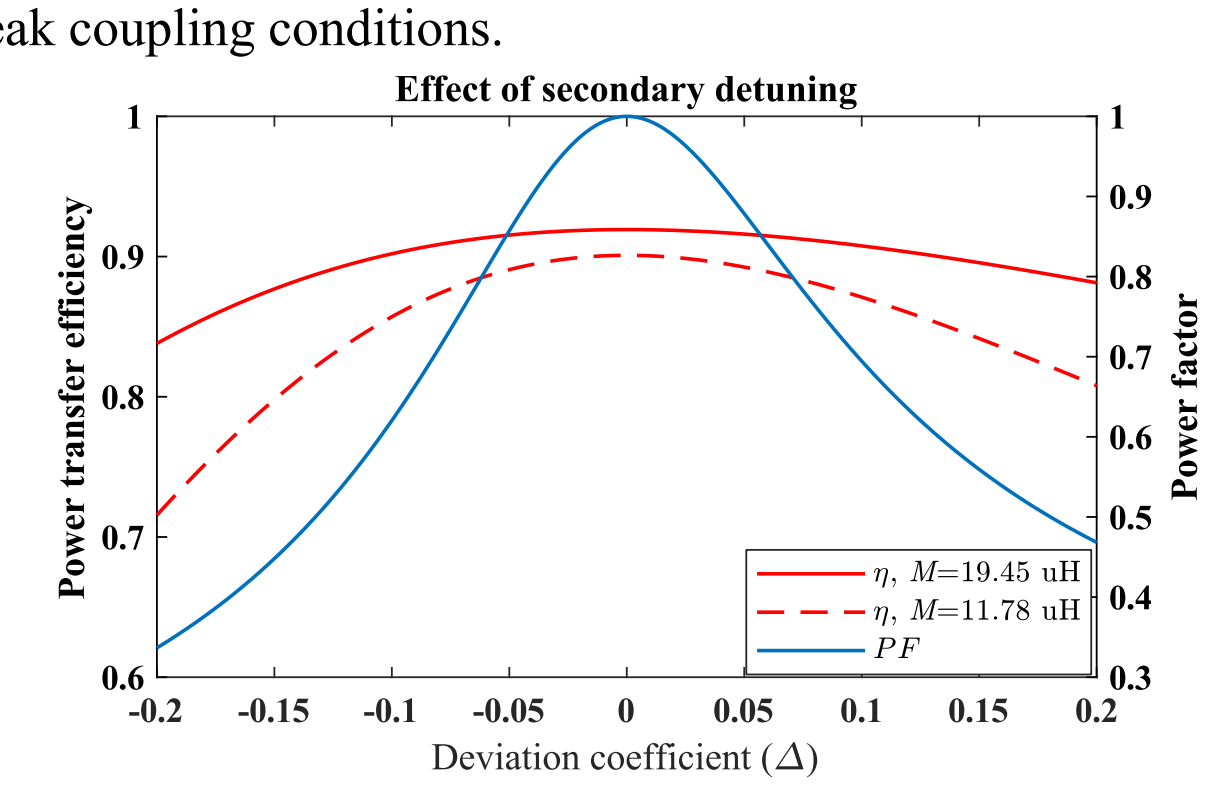

**Fig. 2.** Effect of secondary detuning on system efficiency and power factor.

A comparison of different tuning methods for the detuned WPT system is illustrated in Fig. 3, where the x-axis is the deviation coefficient of the secondary capacitance. Without frequency tuning, efficiency decreases significantly. The operating frequency changes slightly with the primary frequency tuning [8][12], but the efficiency cannot be improved over a wide range. On the other hand, with double-sided frequency tuning, the efficiency can be improved over a wide range [16][17], but this method necessitates additional requirements on receivers. Hence, to simultaneously achieve improved efficiency across a wide range and adaptability of the transmitter, this paper employs the SCC and ZCD circuit on the primary side to swiftly and adaptively track receiver resonant frequency. The proposed scheme ensures a small perturbation of the operating frequency and enables a frequency online tuning capability.

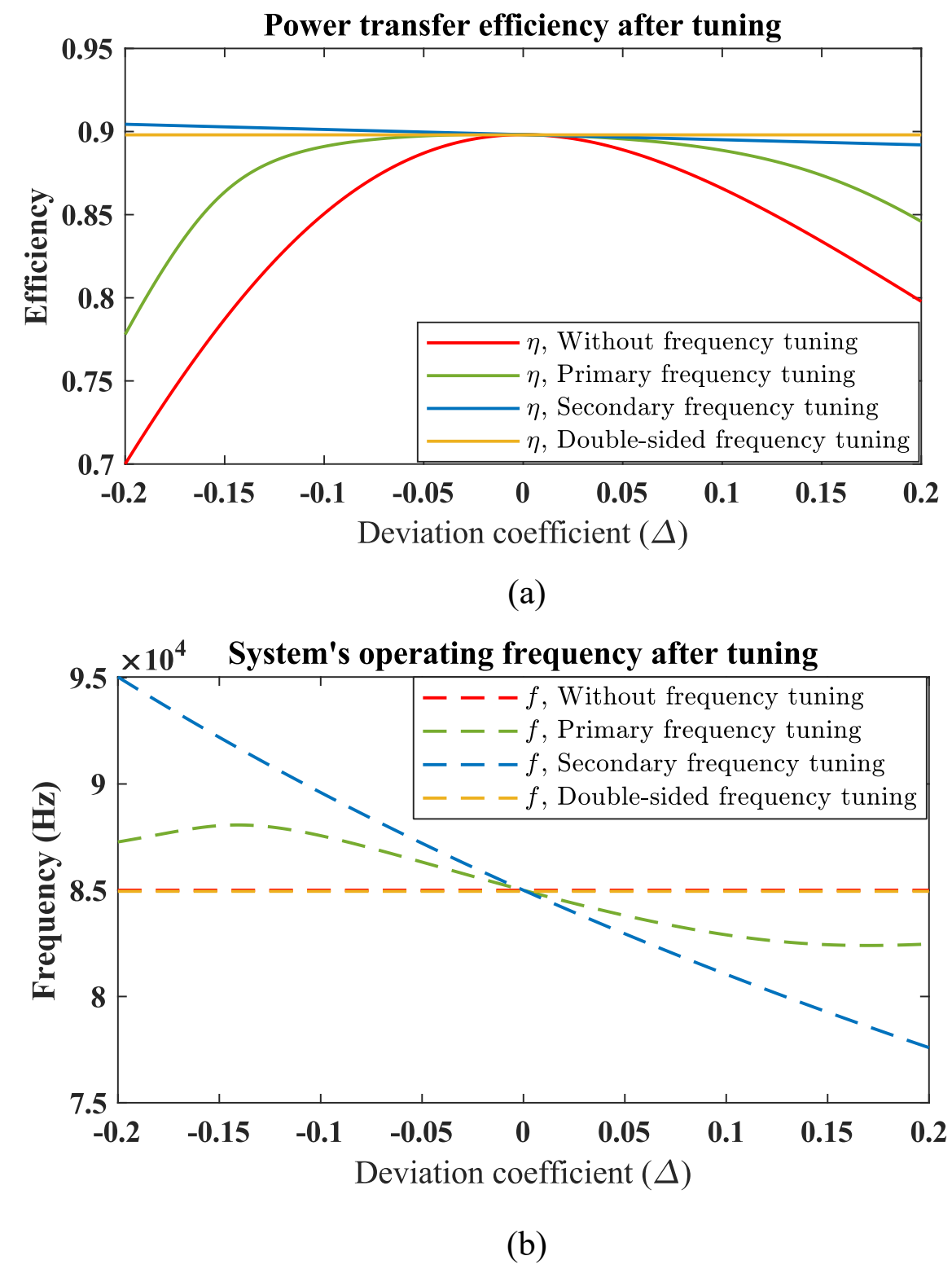

**Fig. 3.** Comparison of different tuning methods for the detuned WPT system caused by variations of the secondary capacitor. (a) Transfer efficiency. (b) Operating frequency.

## III. RECEIVER RESONANT FREQUENCY TRACKING

*A. Identification of receiver resonant frequency*

The following derivations of secondary resonant frequency identification are based on three assumptions.

1) The influence of parasitic resistance $R_p$ and $R_s$ are neglected since, in normal operation range, they are relatively much more minor than $R_e$ or reflected impedance.

2) Primary side reactance $X_p$ is assumed to be 0 under different working frequencies, which is achieved by adjusting the control angle of SCC as depicted in Fig. 4.

3) Given the frequency disturbance's short duration and small magnitude, the equivalent load resistance $R_e$ is assumed to remain constant.

Based on the above assumptions, according to (5), the angle $\theta$ of $Z_{in}$ can be expressed as:

$$\tan\theta = \frac{1/(\omega C_s) - \omega L_s}{R_e} \tag{6}$$

Suppose two known angular frequencies $\omega_m$ and $\omega_n$ are given, and the input impedance angles corresponding to these states are measured; the ratio of the tangents of the two impedance angles can be obtained and defined as $k_{mn}$ in (7).

$$\frac{\tan\theta_m}{\tan\theta_n} = \frac{1/(\omega_m C_s) - \omega_m L_s}{1/(\omega_n C_s) - \omega_n L_s} \triangleq k_{mn} \tag{7}$$

Then, (7) can be transformed as

$$L_s C_s = \frac{(\omega_n - k_{mn}\omega_m)}{\omega_m \omega_n (\omega_m - k_{mn}\omega_n)} \tag{8}$$

By combining equations (1) and (8), the receiver resonant frequency can be obtained as:

$$\omega_s = \sqrt{\frac{\omega_m \omega_n (\omega_m - k_{mn}\omega_n)}{(\omega_n - k_{mn}\omega_m)}} \tag{9}$$

The above derivation demonstrates the high simplicity of the proposed method, as it only requires the addition of a phase detection function to determine the natural resonant frequency of the secondary side swiftly. It does not depend on any parameters of primary or secondary resonators, mutual inductance, and wireless communication.

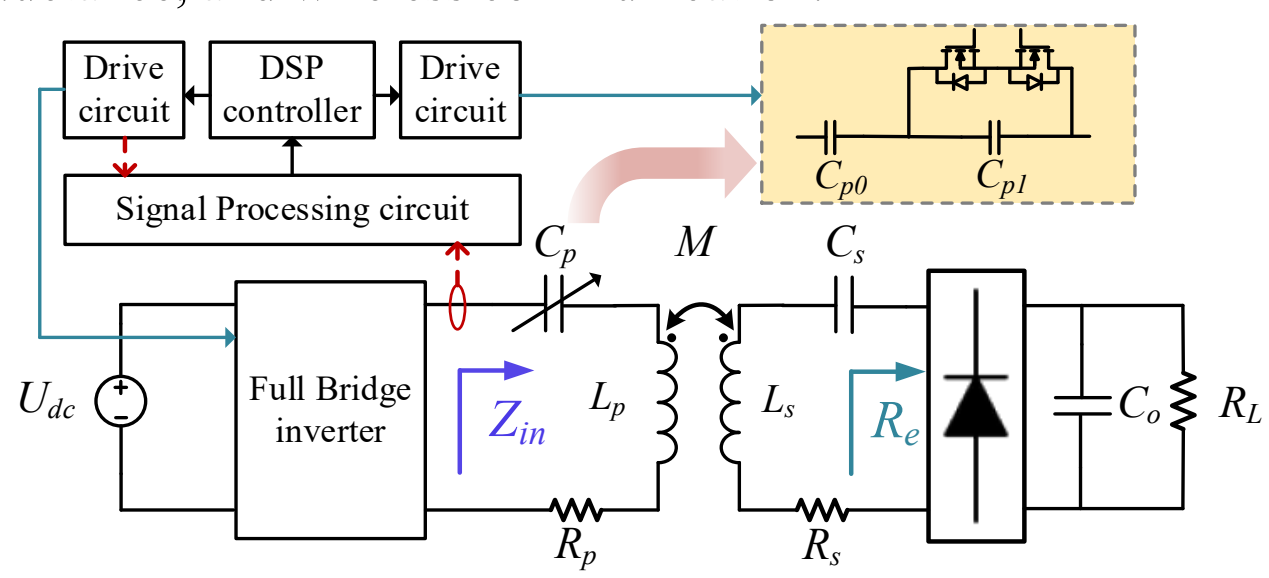

**Fig. 4.** System structure of SS-WPT with secondary resonant frequency identification capability.

## B. Operating Principal of Switch-Controlled Capacitor

The typical structure and waveforms of SCC are shown in Fig. 5, where $i_p$ is the primary resonant current. By alternately activating switch $g_{sc1}$ and $g_{sc2}$, the current $i_p$ can be divided into $i_{sc}$ and $i_{cp1}$, indicating whether $C_{p1}$ is connected in the circuit or short-circuited. In essence, this process allows for control of the equivalent capacitance from $C_{p1}$ to infinite.

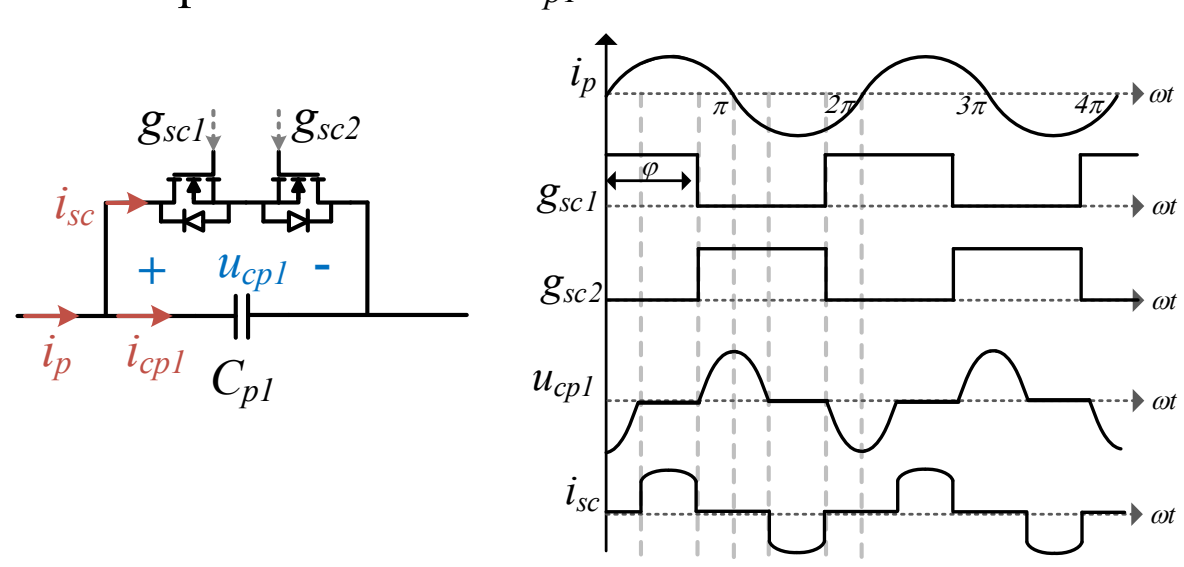

**Fig. 5.** Typical structure and waveforms of SCC.

By performing Fourier analysis on the voltage waveform across the capacitor terminals $u_{cp1}$ at different control angles, the corresponding equivalent capacitance magnitude $C_{sc}$ under the fundamental frequency can be obtained as [20]:

$$C_{sc} = \frac{\pi C_{p1}}{2\pi - 2\varphi + \sin 2\varphi} \tag{10}$$

Connecting the capacitor $C_{p0}$ in series with $C_{sc}$ not only reduces voltage stress but also mitigates the issue of significant variations in equivalent capacitance when the control angle is large [19]. Therefore, the maximum and minimum values of the primary side capacitance $C_p$ can be given by

$$\begin{cases} C_{pmin} = \dfrac{C_{p0} \cdot C_{p1}}{C_{p0} + C_{p1}} \\ C_{pmax} = C_{p0} \end{cases} \tag{11}$$

To ensure that the primary side inductance is compensated at different frequencies, it is necessary to satisfy:

$$[C_{pmin}, C_{pmax}] \supseteq \left[\frac{1}{\omega^2{}_{max} L_p}, \frac{1}{\omega^2{}_{min} L_p}\right] \tag{12}$$

Given the aforementioned conditions, a smaller value of $C_{p0}$ allows it to bear a more significant portion of the voltage across $C_p$, thereby reducing the voltage stress of the SCC.

## C. Analysis of the Proposed Identification Method Under Various Disturbances

Initially, 85kHz is set as the resonant frequency of both the primary and secondary sides. Therefore, the initial selection of two operating frequencies for identification can be designated as 84kHz and 86kHz. This choice of frequencies minimizes the impact on the system output from frequency disturbances while ensuring that phase differences of input impedance remain distinguishable.

As shown in Fig. 6 and Fig. 7, the horizontal axis represents the receiver resonance frequency $f_s$, and $f_{ref}$ indicates the reference line where the identification frequency equals $f_s$. The upper and lower red dotted lines of the $f_{ref}$ curve correspond to the deviation of the identification result by ±1kHz. The purple line $f_{id0}$ represents the identification result without any disturbances. It can be observed that $f_{id0}$ closely matches $f_{ref}$, demonstrating that this method can achieve high-precision identification with different $f_m$ and $f_n$ under the assumptions described earlier. However, during the actual identification process, there may be factors leading to frequency offset during identification, which include:

1) Load disturbance: Due to frequency changes during the identification process, the equivalent load resistance will be subjected to a certain degree of disturbance. This is modelled as a variation in the equivalent load of 0.5 Ω, as shown by the $f_{id1}$ curve in Fig. 6 and Fig. 7.

2) Incomplete compensation of $L_p$ by SCC: Due to factors such as control precision, the value of the equivalent capacitance of SCC deviates from the ideal capacitance corresponding to the complete compensation of inductance. This is illustrated by the $f_{id2}$ curve in Fig. 6 and Fig. 7, whose magnitude of disturbance is about 2% tolerance in the equivalent capacitance of the SCC.

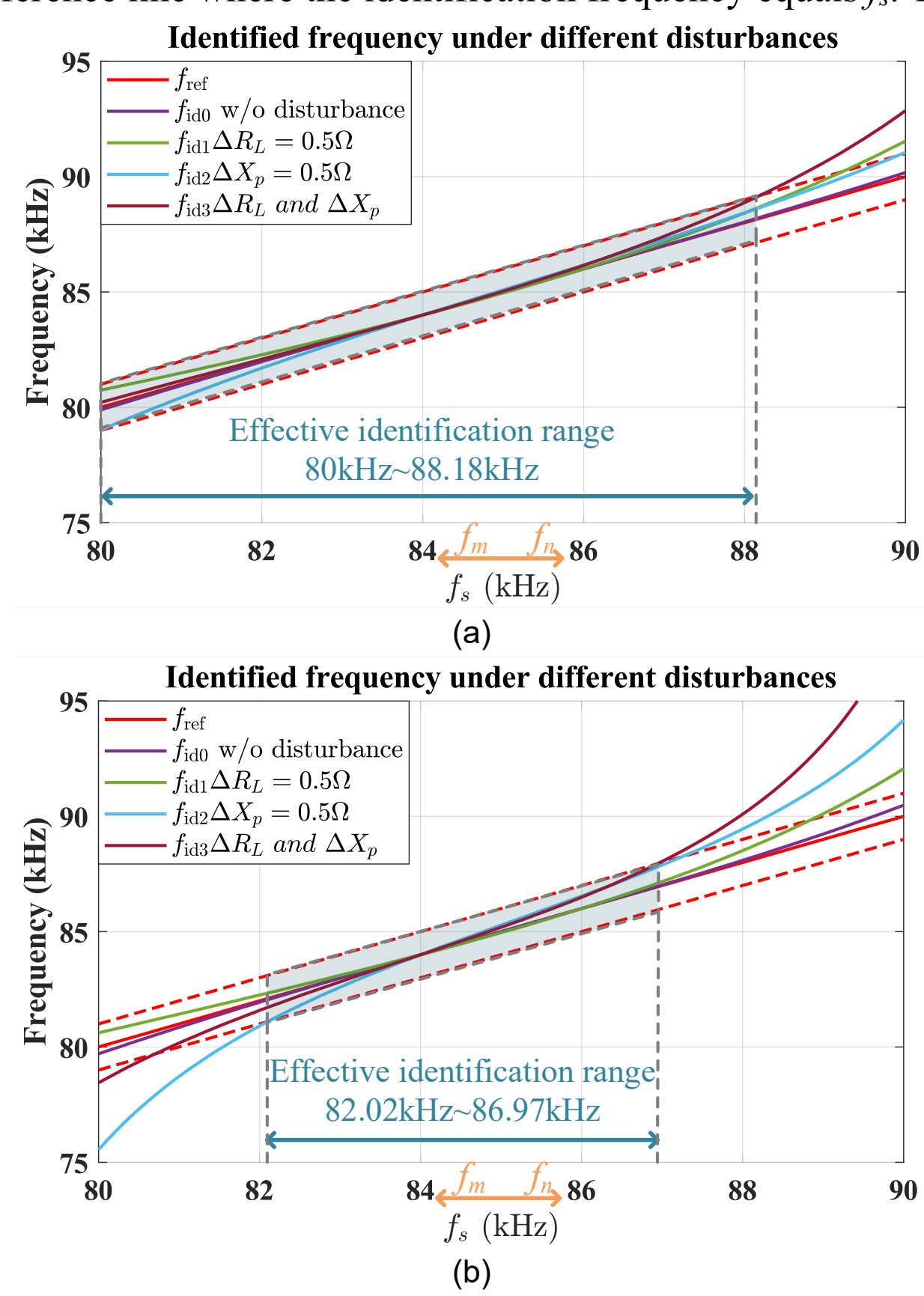

**Fig. 6.** Identified frequency with the variation of receiver resonant frequency with $f_m$ = 84kHz, $f_n$ = 86kHz. (a) without coil misalignment (b) with coil misalignment.

As shown by the curve of $f_{id3}$ in Fig. 6, when load disturbance and incomplete compensation occur simultaneously, the identification results deviate further from the ideal value. When two disturbance frequencies of 84kHz and 86kHz are selected, and there is no coil misalignment, an effective identification range of receiver resonant frequency under various disturbances is from 80kHz to 88.18kHz, considering the error range of ±1kHz. However, when with misalignment, this range narrows to 82.02kHz to 86.97kHz due to decreased mutual inductance and reduced input impedance, leading to a more significant relative error on the input impedance angle under the same disturbance. To proceed, the identification of the $f_s$ at different disturbance frequencies under misaligned conditions shall be analyzed, as shown in Fig. 7.

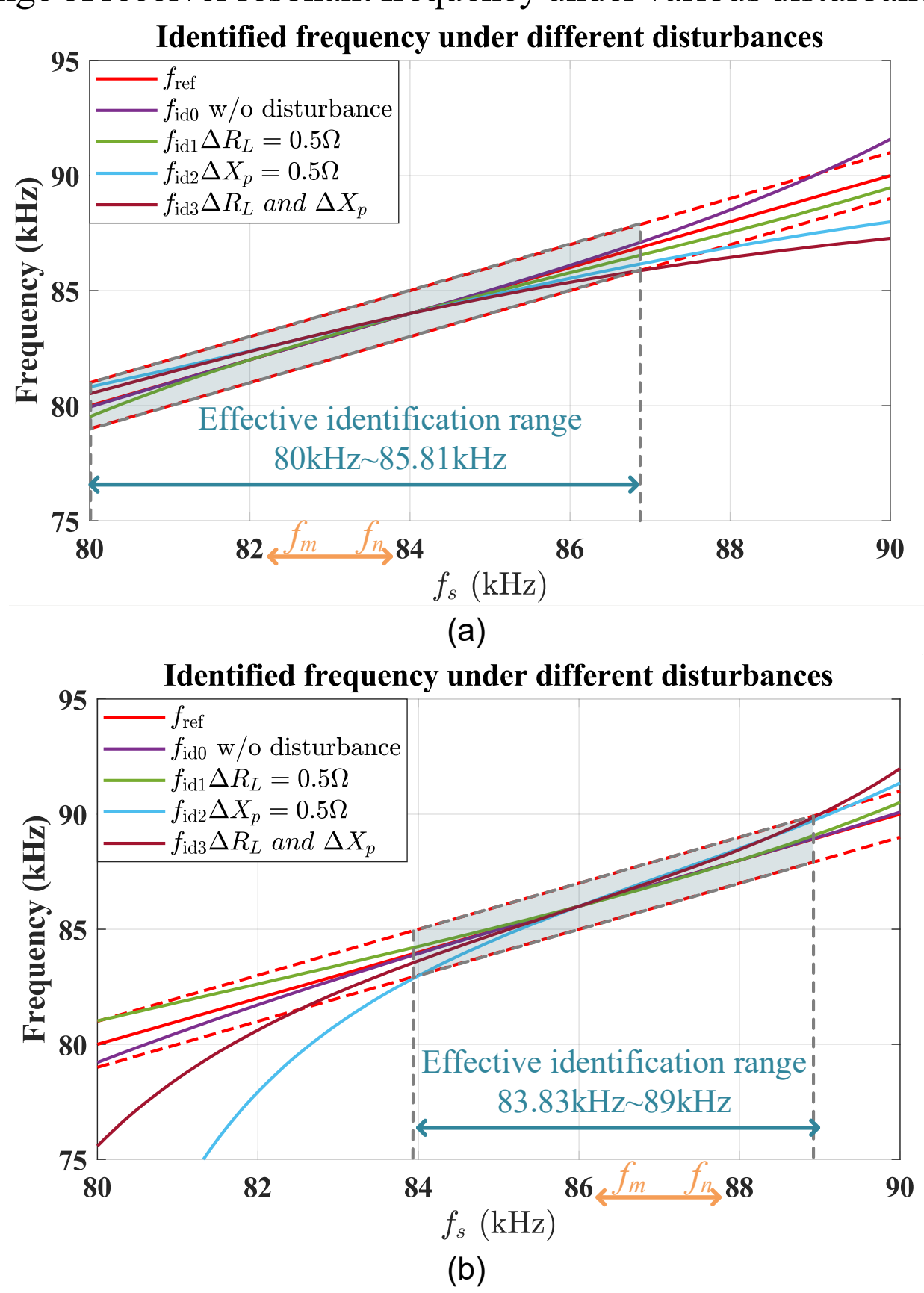

**Fig. 7.** Identified frequency with the variation of secondary resonant frequency under different types of disturbance with coil misalignment. (a) $f_m$ = 84kHz, $f_n$ = 82kHz, (b) $f_m$ = 86kHz, $f_n$ = 88kHz.

Comparing Fig. 6(b) with Fig. 7(a), it is evident that adjusting the disturbance frequencies $f_{m,n}$ from 84 kHz and 86 kHz to 84 kHz and 82 kHz results in a better identification accuracy for the drifted frequency locating in the range of 80~84 kHz. The effective range for identifying $f_s$ also extends from 80 to 85.81 kHz. By comparing Fig. 6(b) with Fig. 7(b), it can be observed that adjusting the disturbance observation frequencies to 86 kHz and 88 kHz results in a significant improvement for the drifting frequency located within the range of 86~90 kHz, with the effective range extending from 83.83 to 89 kHz. Therefore, the identification accuracy can be improved by correctly selecting different disturbance frequencies.

### D. System Configuration and Control Strategy

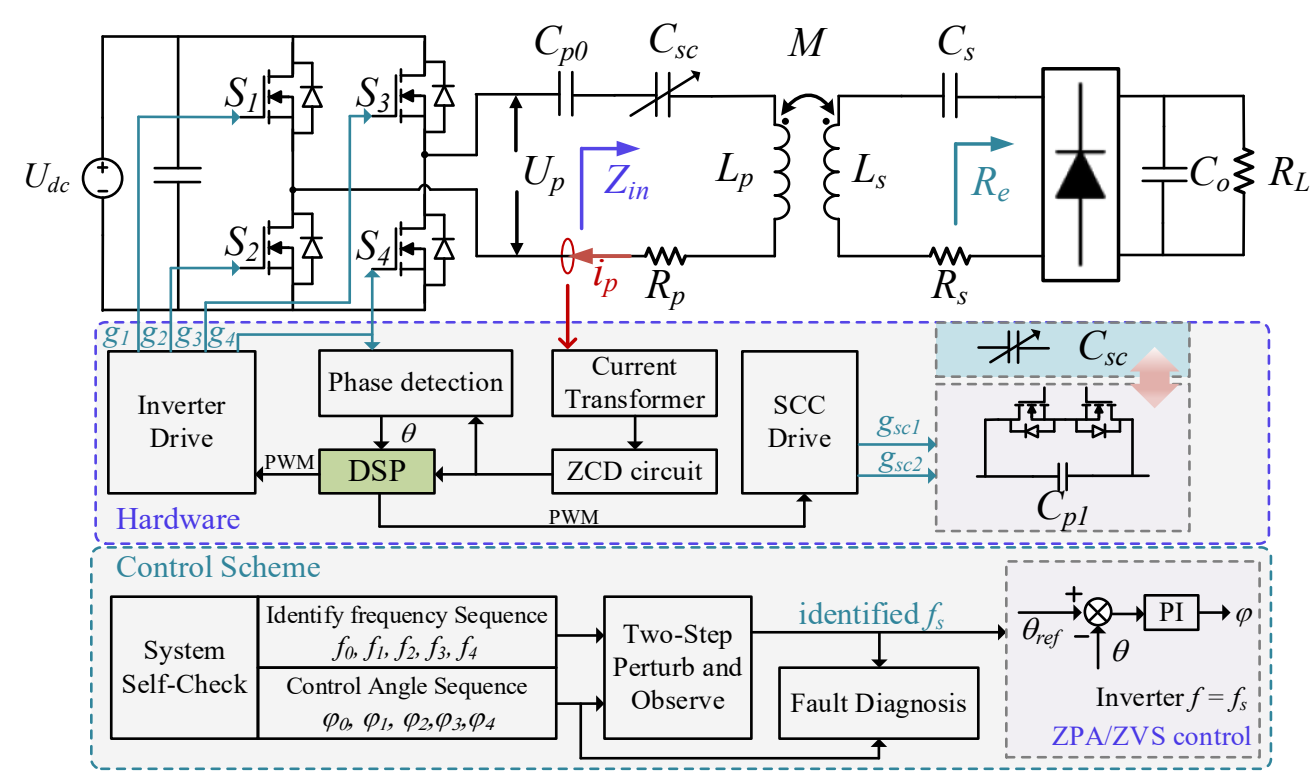

**Fig. 8.** System configuration and control scheme of SS-WPT with adaptive receiver resonant frequency tracking capability.

Fig. 8 illustrates the system configuration and control scheme. The full-bridge inverter provides high-frequency AC power input while the SCC adjusts the capacitance on the primary side. The phase detection of voltage and current is achieved by utilizing the ZCD signal of the current required by the SCC itself and the drive signal of the inverter.

The system operates according to the following workflow:

1) System self-check: When the secondary side is not connected, the input voltage is reduced, and the input current is limited. The inverter frequency is adjusted according to the given frequency sequence for identification, and the SCC control angle sequence for achieving phase alignment of primary side voltage and current is obtained through PI control. In this case, if the obtained control angle sequence cannot achieve the control objective of phase alignment between voltage and current, it is considered that there is a significant deviation in the primary side capacitance, which requires replacement. The fault is reported accordingly.

2) Two-step perturbation and observation: Run the

identification algorithm to determine receiver resonance frequency. Details of the algorithm are presented in Fig. 9.

3) ZPA/ZVS control: To improve the power factor of the primary side operating at $f_s$, PI control is applied to the SCC to ensure that the inverter operates in the ZPA or ZVS state. This reduces the current stress on the primary side and ensures efficient system operation.

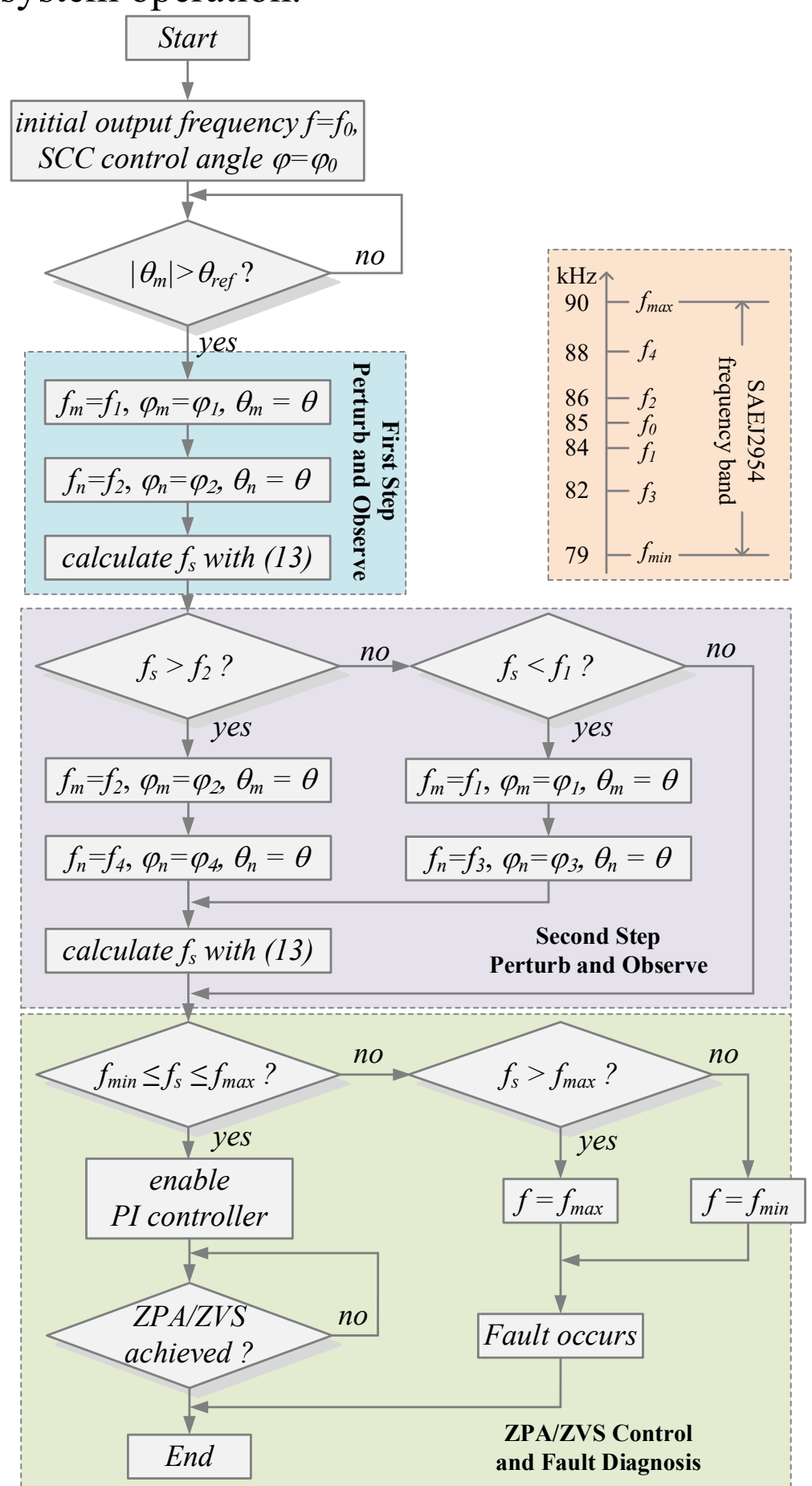

**Fig. 9.** Flowchart of receiver resonant frequency identification and primary compensation strategy.

Fig. 9 illustrates the control flowchart of the system, annotating the various frequencies used ($f_{0,1,2,3,4}$). Initially, the system assumes the inherent resonance frequency of the secondary side as $f_0$. Frequency identification is required when the input impedance angle exceeds a reference value $\theta_{ref}$, indicating the detuning of the secondary side. The identification process is divided into two steps: firstly, perturbation observation is conducted using $f_1$ and $f_2$ to obtain an estimated $f_s$ value. Subsequently, based on this estimate and the magnitudes of $f_1$ and $f_2$, a new set of perturbation frequencies is selected for another observation round, resulting in the final identified $f_s$ value. If this frequency falls within the range specified by the SAE J2954 standard, i.e., from 79 kHz to 90 kHz, then adjust the inverter frequency to this frequency. Otherwise, if severe detuning of the secondary side is detected, report a fault and limit the frequency range accordingly.

## IV. Experimental Verification

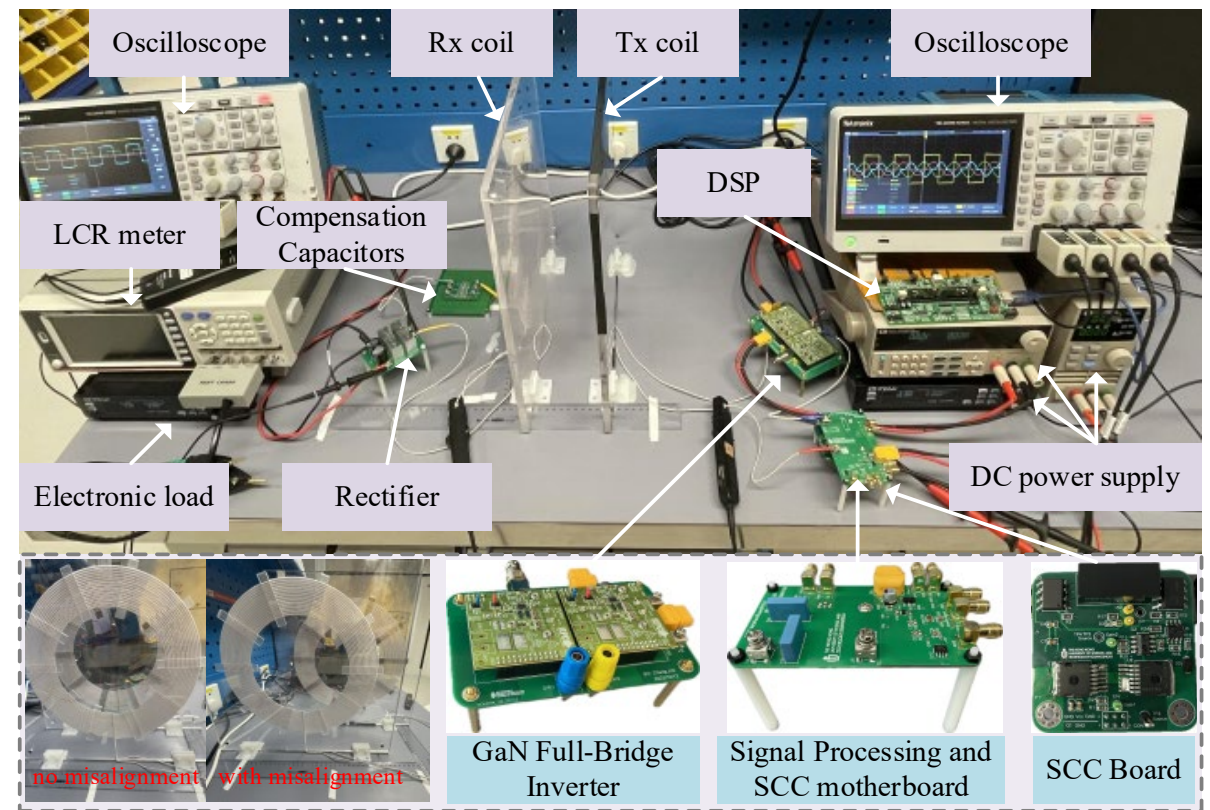

**Fig. 10.** Experimental setup of SS-WPT with secondary resonant frequency identification capability.

The feasibility of the proposed method is demonstrated in this section. As shown in Fig. 10, a 200 W SS-WPT experimental platform is built. EPC 9099 GaN half-bridge modules are used as the inverter. IMBG65R072M1H CoolSiC MOSFETs from Infineon are used for SCC. TMS320F28335 is applied as the digital controller. LCR-8205 from GW Instek measures the system's accurate value of compensation capacitance, self-inductance, and mutual inductance.

To validate the efficacy of the proposed method under various mutual inductance and load conditions, experiments for frequency identification were conducted with and without coil misalignment and under different load scenarios. The essential parameters for experimental validation are outlined in Table III, where system parameters measured under the misalignment condition are denoted in parentheses.

TABLE III
PARAMETERS FOR EXPERIMENT VERIFICATION

| Items | Symbol | Value |
|---|---|---|
| DC Bus Voltage | $U_{dc}$ | 40 V |
| Primary-side Capacitor | $C_{p0}$ | 35.21 nF |
| Primary-side Capacitor | $C_{p1}$ | 98.56 nF |
| Secondary-side Capacitor | $C_s$ | 34.97, 40.79 nF |
| ESR of transmitter | $R_p$ | 0.3 Ω |
| Self-inductance of Tx | $L_p$ | 118.27 (118.30) uH |
| ESR of receiver | $R_s$ | 0.3 Ω |
| Self-inductance of Rx | $L_s$ | 91.95 (91.58) uH |
| Mutual inductance | $M$ | 19.45 (11.78) uH |
| Load resistance | $R_L$ | 4, 8 Ω |

### A. System Self-Check Before Identification

When the receiver is placed at the desired charging position but not connected to the load, a smaller input voltage is used to power the LCR circuit of the primary side and PI control is used to calibrate the SCC control angle at different frequencies so that the output voltage and current of the inverter are in phase, as shown in Fig. 11. There is a slight difference between the SCC control angle obtained by PI control and the control angle obtained according to (10). This may be due to factors such as the tolerance of the capacitor itself and the delay of the SCC drive circuit.

### *B. Frequency Identification*

Table IV presents the settings for different cases during the experiment, designed to individually assess the effectiveness of frequency identification across varying frequencies, mutual inductances, and loads.

TABLE IV
EXPERIMENT CONDITIONS

| Cases | $C_s$ | $f_s$(kHz) | Misalignment | $R_L$ | Control |
|---|---|---|---|---|---|
| I | 40.79 | 82.178 | Without | 8 Ω | ZPA |
| II | 34.97 | 88.756 | Without | 8 Ω | ZVS |
| III | 34.97 | 88.934 | With | 8 Ω | ZVS |
| IV | 34.97 | 88.934 | With | 4 Ω | ZPA |

Fig. 12 provides the system waveforms before and after frequency tuning when the coils are aligned. Case I and II have the same load but different receiver resonant frequencies in this scenario. The identification errors are 0.382 kHz and -0.196 kHz, respectively. At the same time, ZPA and ZVS control are realized, respectively.

Similarly, the model can successfully identify the receiver resonant frequency without adjusting to model parameters when there are changes in coil position and load, as depicted in Fig. 13 for Case III and IV. Moreover, in instances of misalignment, the increased stress on the primary-side current due to detuning on the secondary side is evident before the tuning process in Case III. Hence, the rapid, precise, and real-time monitoring of the receiver resonant frequency is essential for ensuring system stability. Case III and IV achieve ZVS or ZPA, with estimated errors of -0.354 and -0.624 kHz, respectively. As demonstrated in Fig. 12 and Fig. 13, a 90-degree phase shift between the primary and secondary side current is observed, confirming the accuracy of the frequency identification algorithm.

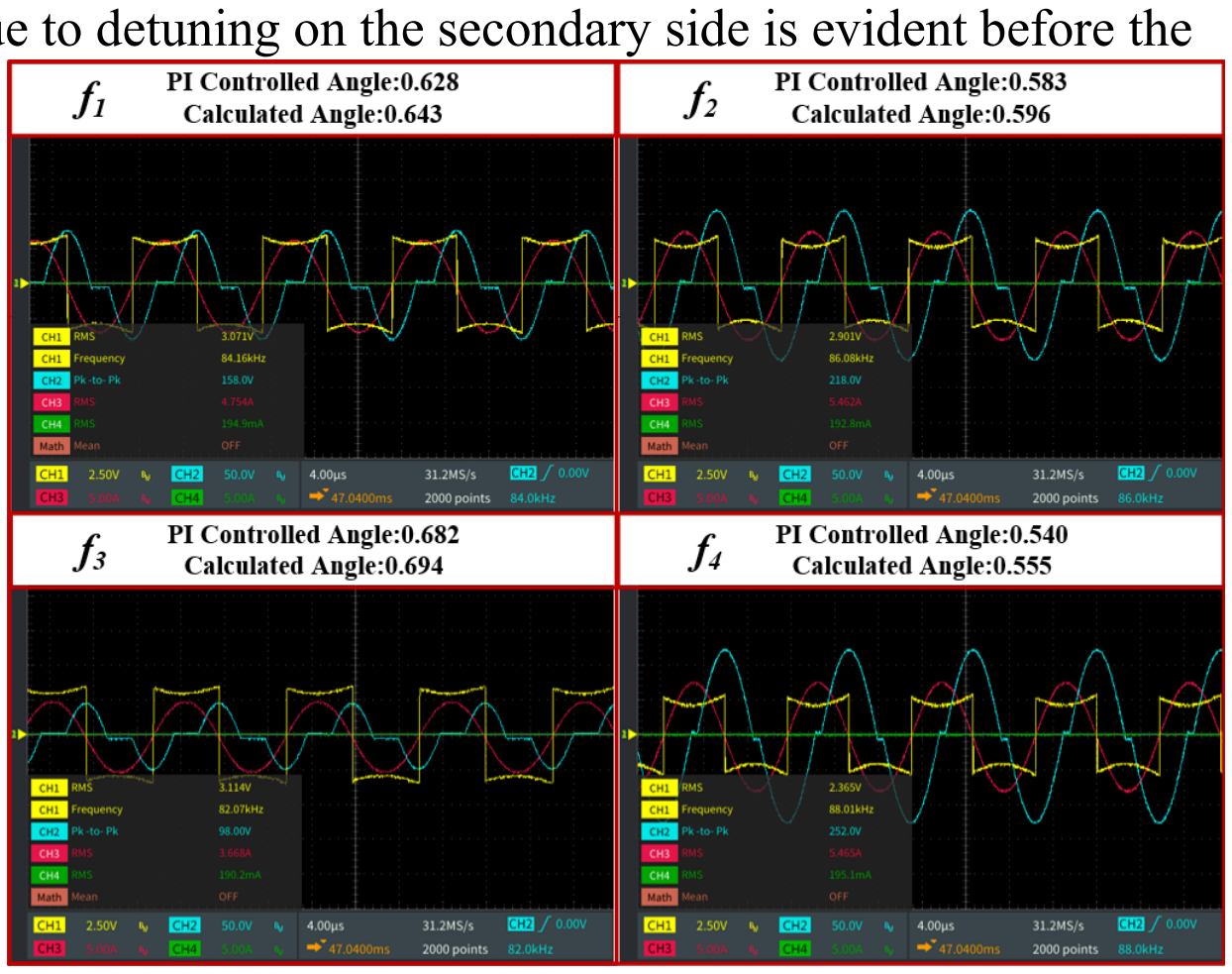

**Fig. 11.** Experimental waveforms of system self-check process.

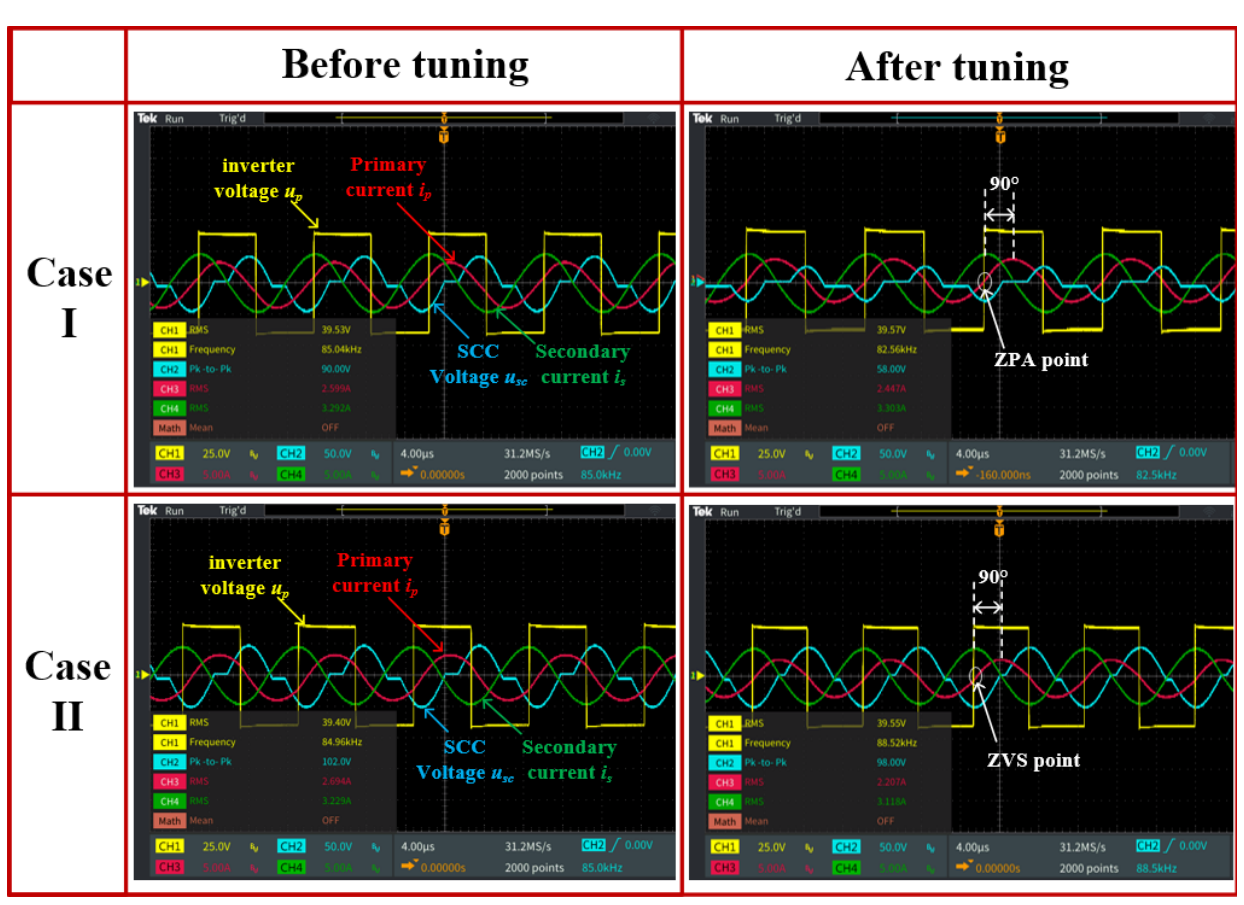

**Fig. 12.** Experimental waveforms before and after the tuning process without misalignment.

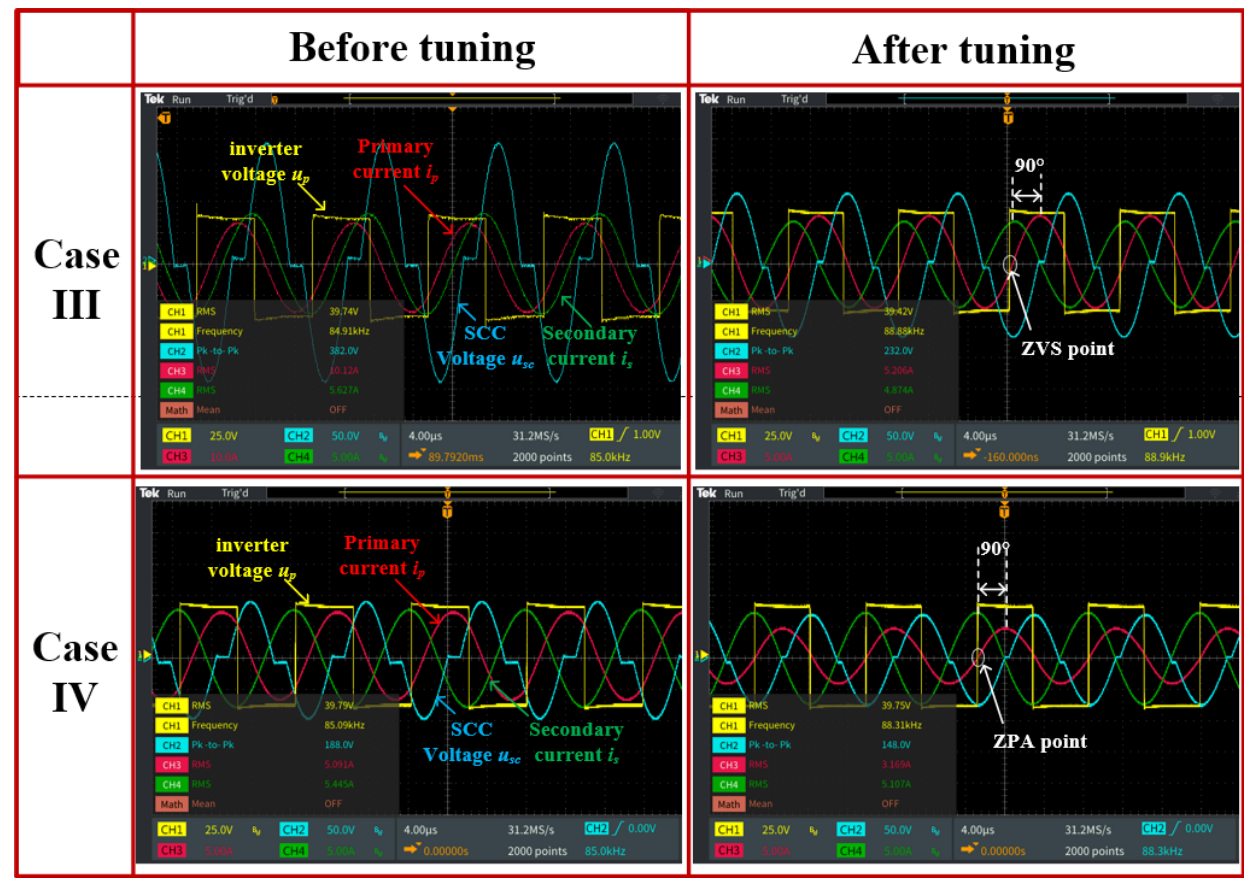

**Fig. 13.** Experimental waveforms before and after the tuning process with misalignment.

### *C. Response time*

Fig. 14 and Fig. 15, respectively, depict the transient changes on the primary and secondary sides of the SS-WPT system before and after using the proposed frequency tuning scheme. Both figures demonstrate that the two-step perturb-and-observe algorithm accomplishes identification within 1 ms, swiftly adjusting the output to the receiver resonance frequency. In comparison to the previously fastest decoupling control-based approach [17], the speed doubles, and there is no longer a requirement for complex decoupling transformer design.

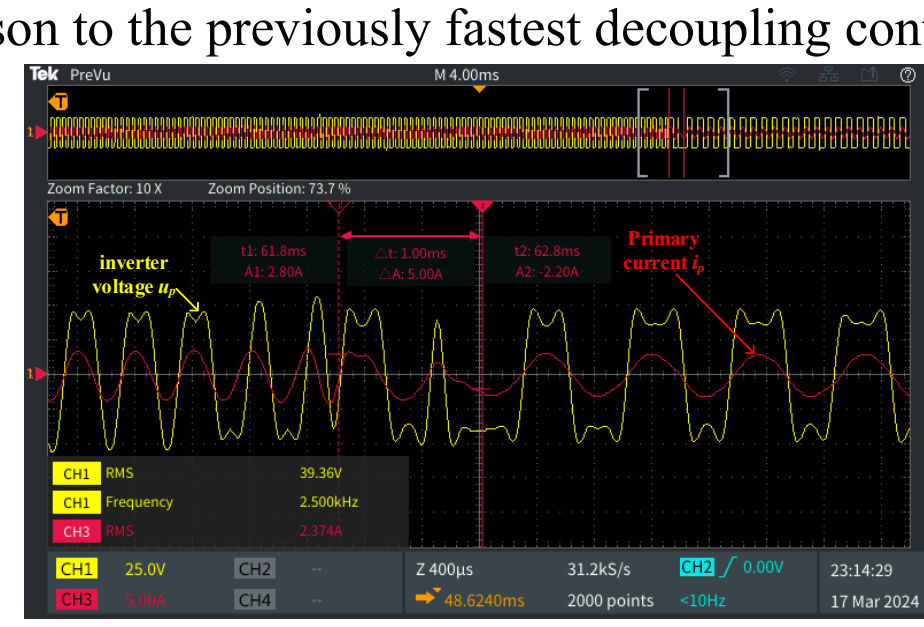

**Fig. 14.** Transient waveforms of inverter output voltage and current before and after frequency tuning.

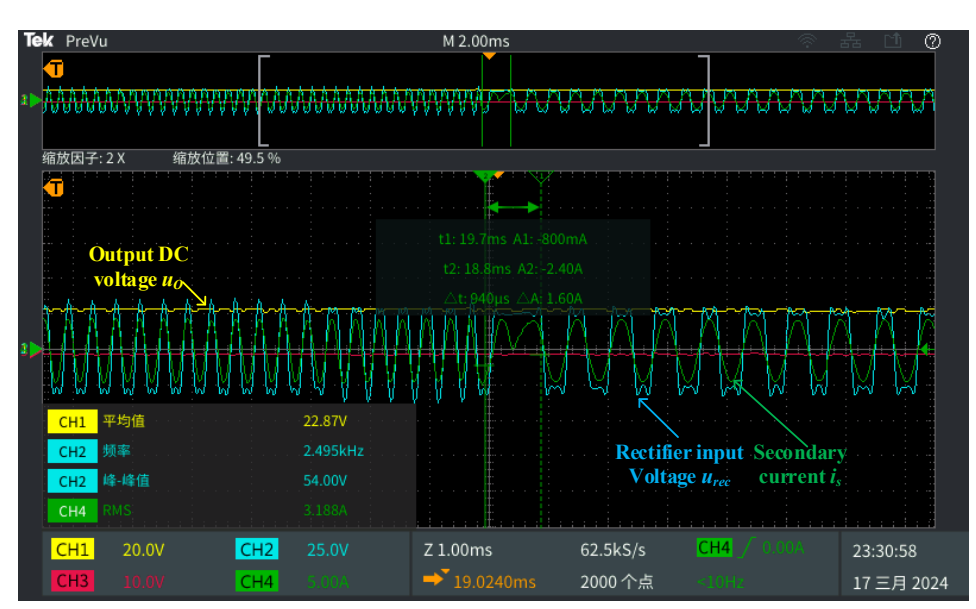

**Fig. 15.** Transient waveforms of rectifier input voltage and current before and after frequency tuning.

### *D. Efficiency Comparison*

As shown in Fig. 16, the changes in DC output power and efficiency under different cases before and after frequency tuning are illustrated. In cases without misalignment, where there is relatively high mutual inductance, the improvement in system efficiency is not significant, approximately 1% to 3%. However, the input current stress is effectively reduced, as seen in Fig. 12, while maintaining a relatively stable output power. In the misaligned cases, the system's efficiency is improved by 9% after tuning. It also enhances the power factor on the input side, as shown in Fig. 13. Since this study focuses on the frequency range specified by the SAE J2954 standard for identification and experimentation, the frequency deviation range is not too extensive. In practical applications, the aging and faults of resonant capacitors may significantly mitigate the system's performance.

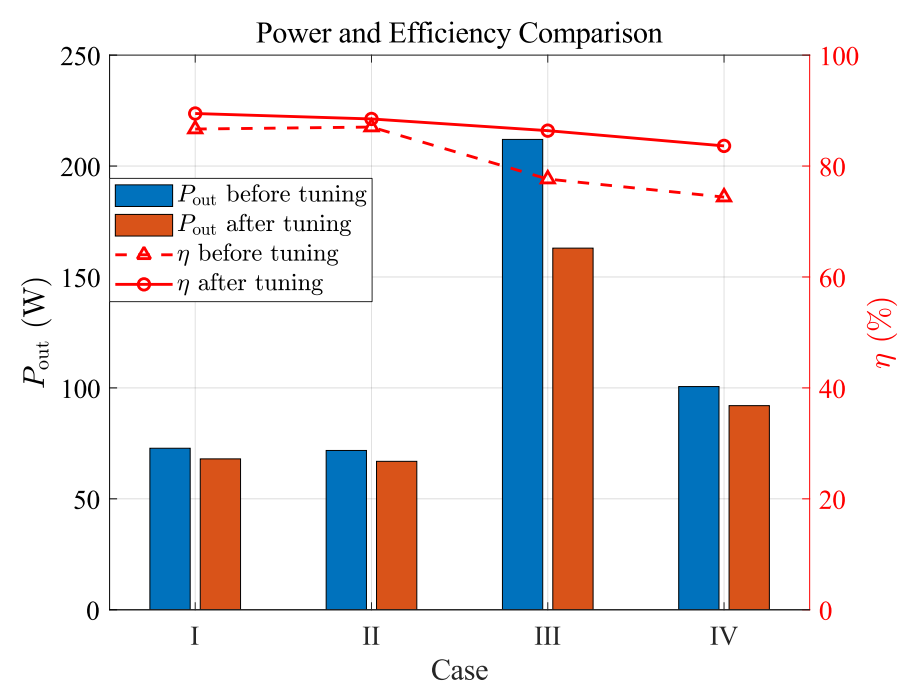

**Fig. 16.** Output power and system efficiency of different experimental cases before and after tuning.

## V. Conclusion

This article introduces an adaptive receiver resonant frequency identification and tracking method utilizing only primary-side information. The approach leverages SCC and the current ZCD circuit to cancel primary reactance and maintain ZVS operation. Notably, this method achieves frequency identification within 1 ms and with a maximum identification error of ±0.7 kHz without requiring the parameters of the original primary and secondary sides or wireless communication. It can effectively detect system detuning faults and improve the transmitter's adaptability in SS-WPT systems to different receivers that comply with the SAE J2954 standard.